\catcode`\@=11					% To make protected \def's

%************************************************************
%*
%*		Font set-up
%*
%************************************************************

%************** 5-point fonts *******************************

\font\fiverm=cmr5				% roman
\font\fivemi=cmmi5				% math italic
\font\fivesy=cmsy5				% math symbols
\font\fivebf=cmbx5				% bold face

\skewchar\fivemi='177
\skewchar\fivesy='60

%************** 6-point fonts *******************************

\font\sixrm=cmr6				% roman
\font\sixi=cmmi6				% math italic
\font\sixsy=cmsy6				% math symbols
\font\sixbf=cmbx6				% bold face

\skewchar\sixi='177
\skewchar\sixsy='60

%************** 7-point fonts *******************************

\font\sevenrm=cmr7				% roman
\font\seveni=cmmi7				% math italic
\font\sevensy=cmsy7				% math symbols
\font\sevenit=cmti7				% italic
\font\sevenbf=cmbx7				% bold face

\skewchar\seveni='177
\skewchar\sevensy='60

%************** 8-point fonts *******************************

\font\eightrm=cmr8				% roman
\font\eighti=cmmi8				% math italic
\font\eightsy=cmsy8				% math symbols
\font\eightit=cmti8				% italic
				% slanted
\font\eightbf=cmbx8				% bold face
				% typewriter
				% sans serif

\skewchar\eighti='177
\skewchar\eightsy='60

%************** 9-point fonts *******************************

\font\ninei=cmmi9
\font\ninesy=cmsy9

\skewchar\ninei='177
\skewchar\ninesy='60

%************** 10-point fonts ******************************

\font\tenrm=cmr10				% roman
\font\teni=cmmi10				% math italic
\font\tensy=cmsy10				% math symbols
\font\tenex=cmex10				% math extension
\font\tenit=cmti10				% italic
\font\tensl=cmsl10				% slanted
\font\tenbf=cmbx10				% bold face
\font\tentt=cmtt10				% typewriter
\font\tenss=cmss10				% sans serif
\font\tensc=cmcsc10				% small caps
\font\tenbi=cmmib10				% bold math

\skewchar\teni='177
\skewchar\tenbi='177
\skewchar\tensy='60

\def\tenpoint{\ifmmode\err@badsizechange\else
	\textfont0=\tenrm \scriptfont0=\sevenrm \scriptscriptfont0=\fiverm
	\textfont1=\teni  \scriptfont1=\seveni  \scriptscriptfont1=\fivemi
	\textfont2=\tensy \scriptfont2=\sevensy \scriptscriptfont2=\fivesy
	\textfont3=\tenex \scriptfont3=\tenex   \scriptscriptfont3=\tenex
	\textfont4=\tenit \scriptfont4=\sevenit \scriptscriptfont4=\sevenit
	\textfont5=\tensl
	\textfont6=\tenbf \scriptfont6=\sevenbf \scriptscriptfont6=\fivebf
	\textfont7=\tentt
	\textfont8=\tenbi \scriptfont8=\seveni  \scriptscriptfont8=\fivemi
	\def\rm{\tenrm\fam=0 }%
	\def\it{\tenit\fam=4 }%
	\def\sl{\tensl\fam=5 }%
	\def\bf{\tenbf\fam=6 }%
	\def\tt{\tentt\fam=7 }%
	\def\ss{\tenss}%
	\def\sc{\tensc}%
	\def\bmit{\fam=8 }%
	\rm\setparameters\setbaselines\fi}

%************** 12-point fonts ******************************

\font\twelverm=cmr12				% roman
\font\twelvei=cmmi12				% math italic
\font\twelvesy=cmsy10	scaled\magstep1		% math symbols
\font\twelveex=cmex10	scaled\magstep1		% math extension
\font\twelveit=cmti12				% italic
\font\twelvesl=cmsl12				% slanted
\font\twelvebf=cmbx12				% bold face
\font\twelvett=cmtt12				% typewriter
\font\twelvess=cmss12				% sans serif
\font\twelvesc=cmcsc10	scaled\magstep1		% small caps
\font\twelvebi=cmmib10	scaled\magstep1		% bold math

\skewchar\twelvei='177
\skewchar\twelvebi='177
\skewchar\twelvesy='60

\def\twelvepoint{\ifmmode\err@badsizechange\else
	\textfont0=\twelverm \scriptfont0=\eightrm \scriptscriptfont0=\sixrm
	\textfont1=\twelvei  \scriptfont1=\eighti  \scriptscriptfont1=\sixi
	\textfont2=\twelvesy \scriptfont2=\eightsy \scriptscriptfont2=\sixsy
	\textfont3=\twelveex \scriptfont3=\tenex   \scriptscriptfont3=\tenex
	\textfont4=\twelveit \scriptfont4=\eightit \scriptscriptfont4=\sevenit
	\textfont5=\twelvesl
	\textfont6=\twelvebf \scriptfont6=\eightbf \scriptscriptfont6=\sixbf
	\textfont7=\twelvett
	\textfont8=\twelvebi \scriptfont8=\eighti  \scriptscriptfont8=\sixi
	\def\rm{\twelverm\fam=0 }%
	\def\it{\twelveit\fam=4 }%
	\def\sl{\twelvesl\fam=5 }%
	\def\bf{\twelvebf\fam=6 }%
	\def\tt{\twelvett\fam=7 }%
	\def\ss{\twelvess}%
	\def\sc{\twelvesc}%
	\def\bmit{\fam=8 }%
	\rm\setparameters\setbaselines\fi}

%************** 14-point fonts ******************************

\font\fourteenrm=cmr12	scaled\magstep1		% roman
\font\fourteeni=cmmi12	scaled\magstep1		% math italic
\font\fourteensy=cmsy10	scaled\magstep2		% math symbols
\font\fourteenex=cmex10	scaled\magstep2		% math extension
\font\fourteenit=cmti12	scaled\magstep1		% italic
\font\fourteensl=cmsl12	scaled\magstep1		% slanted
\font\fourteenbf=cmbx12	scaled\magstep1		% bold face
\font\fourteentt=cmtt12	scaled\magstep1		% typewriter
\font\fourteenss=cmss12	scaled\magstep1		% sans serif
\font\fourteensc=cmcsc10 scaled\magstep2	% small caps
\font\fourteenbi=cmmib10 scaled\magstep2	% bold math

\skewchar\fourteeni='177
\skewchar\fourteenbi='177
\skewchar\fourteensy='60

\def\fourteenpoint{\ifmmode\err@badsizechange\else
	\textfont0=\fourteenrm \scriptfont0=\tenrm \scriptscriptfont0=\sevenrm
	\textfont1=\fourteeni  \scriptfont1=\teni  \scriptscriptfont1=\seveni
	\textfont2=\fourteensy \scriptfont2=\tensy \scriptscriptfont2=\sevensy
	\textfont3=\fourteenex \scriptfont3=\tenex \scriptscriptfont3=\tenex
	\textfont4=\fourteenit \scriptfont4=\tenit \scriptscriptfont4=\sevenit
	\textfont5=\fourteensl
	\textfont6=\fourteenbf \scriptfont6=\tenbf \scriptscriptfont6=\sevenbf
	\textfont7=\fourteentt
	\textfont8=\fourteenbi \scriptfont8=\tenbi \scriptscriptfont8=\seveni
	\def\rm{\fourteenrm\fam=0 }%
	\def\it{\fourteenit\fam=4 }%
	\def\sl{\fourteensl\fam=5 }%
	\def\bf{\fourteenbf\fam=6 }%
	\def\tt{\fourteentt\fam=7}%
	\def\ss{\fourteenss}%
	\def\sc{\fourteensc}%
	\def\bmit{\fam=8 }%
	\rm\setparameters\setbaselines\fi}

%************** Miscellaneous big fonts *********************

		% roman
		% bold face

%************************************************************
%*
%*		Parameter initialization
%*
%************************************************************

\newdimen\rp@
\newcount\@basestretchnum
\newskip\@baseskip
\newskip\headskip
\newskip\footskip

% Routine to set page parameters

\def\setparameters{\rp@=.1em
	\headskip=24\rp@
	\footskip=\headskip
	\delimitershortfall=5\rp@
	\nulldelimiterspace=1.2\rp@
	\scriptspace=0.5\rp@
	\abovedisplayskip=10\rp@ plus3\rp@ minus5\rp@
	\belowdisplayskip=10\rp@ plus3\rp@ minus5\rp@
	\abovedisplayshortskip=5\rp@ plus2\rp@ minus4\rp@
	\belowdisplayshortskip=10\rp@ plus3\rp@ minus5\rp@
	\normallineskip=\rp@
	\lineskip=\normallineskip
	\normallineskiplimit=0pt
	\lineskiplimit=\normallineskiplimit
	\jot=3\rp@
	\setbox0=\hbox{\the\textfont3 B}\p@renwd=\wd0
	\skip\footins=12\rp@ plus3\rp@ minus3\rp@
	\skip\topins=0pt plus0pt minus0pt}

% Special routine to scale \baselineskip

\def\setbaselines{\maxdepth=4\rp@\baselinestretch=\@basestretchnum}

% The \baselinestretch command

\def\baselinestretch{\afterassignment\@basestretch\@basestretchnum}
\def\@basestretch{%
	\@baseskip=12\rp@ \divide\@baseskip by1000
	\normalbaselineskip=\@basestretchnum\@baseskip
	\baselineskip=\normalbaselineskip
	\bigskipamount=\the\baselineskip
		plus.25\baselineskip minus.25\baselineskip
	\medskipamount=.5\baselineskip
		plus.125\baselineskip minus.125\baselineskip
	\smallskipamount=.25\baselineskip
		plus.0625\baselineskip minus.0625\baselineskip
	\setbox\strutbox=\hbox{\vrule height.708\baselineskip
		depth.292\baselineskip width0pt }}

%************************************************************
%*
%*		Modifications to PLAIN.TEX
%*
%************************************************************

% Modifications to PLAIN routines to handle scaling of page parameters

\def\makeheadline{\vbox to0pt{\baselinestretch=1000
	\vskip-\headskip \vskip1.5pt
	\line{\vbox to\ht\strutbox{}\the\headline}\vss}\nointerlineskip}

\def\makefootline{\baselineskip=\footskip\line{\the\footline}}

\def\big#1{{\hbox{$\left#1\vbox to8.5\rp@ {}\right.\n@space$}}}
\def\Big#1{{\hbox{$\left#1\vbox to11.5\rp@ {}\right.\n@space$}}}
\def\bigg#1{{\hbox{$\left#1\vbox to14.5\rp@ {}\right.\n@space$}}}
\def\Bigg#1{{\hbox{$\left#1\vbox to17.5\rp@ {}\right.\n@space$}}}

% Modifications to PLAIN to handle bold math

\mathchardef\alpha="710B
\mathchardef\beta="710C
\mathchardef\gamma="710D
\mathchardef\delta="710E
\mathchardef\epsilon="710F
\mathchardef\zeta="7110
\mathchardef\eta="7111
\mathchardef\theta="7112
\mathchardef\iota="7113
\mathchardef\kappa="7114
\mathchardef\lambda="7115
\mathchardef\mu="7116
\mathchardef\nu="7117
\mathchardef\xi="7118
\mathchardef\pi="7119
\mathchardef\rho="711A
\mathchardef\sigma="711B
\mathchardef\tau="711C
\mathchardef\upsilon="711D
\mathchardef\phi="711E
\mathchardef\chi="711F
\mathchardef\psi="7120
\mathchardef\omega="7121
\mathchardef\varepsilon="7122
\mathchardef\vartheta="7123
\mathchardef\varpi="7124
\mathchardef\varrho="7125
\mathchardef\varsigma="7126
\mathchardef\varphi="7127
\mathchardef\imath="717B
\mathchardef\jmath="717C
\mathchardef\ell="7160
\mathchardef\wp="717D
\mathchardef\partial="7140
\mathchardef\flat="715B
\mathchardef\natural="715C
\mathchardef\sharp="715D

%************************************************************
%*
%*		Initialization
%*
%************************************************************

\def\err@badsizechange{%
	\immediate\write16{--> Size change not allowed in math mode, ignored}}

\baselinestretch=1000
\tenpoint

\catcode`\@=12					% Restore @ sign
% Routine to guarantee that this file is input only once
\catcode`\@=11
\expandafter\ifx\csname @iasmacros\endcsname\relax
	\global\let\@iasmacros=\par
\else	\immediate\write16{}
	\immediate\write16{Warning:}
	\immediate\write16{You have tried to input iasmacros more than once.}
	\immediate\write16{}
	\endinput
\fi
\catcode`\@=12

% Set up font size commands and \baselinestretch command
%\input iasfonts

% Some alternative font names

% Simple spacing commands
\def\singlespace{\baselineskip=\normalbaselineskip}
\def\halfspace{\baselineskip=1.5\normalbaselineskip}
\def\doublespace{\baselineskip=2\normalbaselineskip}

% Macros for references and abstracts

\def\AB{\bigskip\parindent=40pt
        \centerline{\bf ABSTRACT}\medskip\halfspace\narrower}
\def\AE{\bigskip\nonarrower\doublespace}
\def\nonarrower{\advance\leftskip by-\parindent
	\advance\rightskip by-\parindent}

% Useful commands

\def\boxit#1{\vbox{\hrule\hbox{\vrule\kern3pt
	\vbox{\kern3pt#1\kern3pt}\kern3pt\vrule}\hrule}}

% Special symbols
\def\hence{\leavevmode\hbox{\bf .\raise5.5pt\hbox{.}.} }

\def\dalemb#1#2{{\vbox{\hrule height.#2pt
	\hbox{\vrule width.#2pt height#1pt \kern#1pt \vrule width.#2pt}
	\hrule height.#2pt}}}
\def\gtorder{\mathrel{\raise.3ex\hbox{$>$}\mkern-14mu
             \lower0.6ex\hbox{$\sim$}}}
\def\ltorder{\mathrel{\raise.3ex\hbox{$<$}\mkern-14mu
             \lower0.6ex\hbox{$\sim$}}}

% For twoup output
\newdimen\fullhsize
\newbox\leftcolumn
\def\twoup{\hoffset=-.5in \voffset=-.25in
  \hsize=4.75in \fullhsize=10in \vsize=6.9in
  \def\fullline{\hbox to\fullhsize}
  \let\lr=L
  \output={\if L\lr
        \global\setbox\leftcolumn=\columnbox\global\let\lr=R \advancepageno
      \else \doubleformat \global\let\lr=L\fi
    \ifnum\outputpenalty>-20000 \else\dosupereject\fi}
  \def\doubleformat{\shipout\vbox{
    \fullline{\box\leftcolumn\hfil\columnbox}\advancepageno}}
  \def\columnbox{\leftline{\vbox{\makeheadline\pagebody\makefootline}}}
  \tolerance=1000 }
\twelvepoint
\doublespace
{\nopagenumbers{
\rightline{~~~Jan., 2002}
\bigskip\bigskip
\centerline{\fourteenpoint Should $E_8$ SUSY Yang-Mills be Reconsidered}
\centerline{\fourteenpoint as a  Family Unification Model?}
\medskip
\centerline{\it Stephen L. Adler
%{\singlespace { Research supported in part by the Department of Energy 
%under Grant No.
%DE-FG02-90ER40542.}}
}
\centerline{\bf Institute for Advanced Study}
\centerline{\bf Princeton, NJ 08540}
\medskip
%\leftline{{\it Short title:} short title}
\bigskip\bigskip
\leftline{\it Send correspondence to:}
\medskip
{\singlespace\leftline{Stephen L. Adler}
\leftline{Institute for Advanced Study}
\leftline{Einstein Drive, Princeton, NJ 08540}
\leftline{Phone 609-734-8051; FAX 609-924-8399; email adler@ias.
edu}}
\bigskip\bigskip
}}
\vfill\eject
\pageno=2
\AB
We review earlier proposals for $E_8$ family unification, and discuss  
why recent work of Kovner and Shifman on condensates in supersymmetric  
Yang-Mills theories suggests 
the reconsideration of $E_8$ supersymmetric Yang-Mills as a family 
unification theory.  
\AE
\bigskip\bigskip
\vfill\eject
\pageno=3
One of the outstanding mysteries of the current standard model is the 
triple repetition of fundamental fermions.  Many different ideas [1] have 
been proposed to explain why there are three (or in some models more) 
families; here we focus on the possibility, already addressed in the earlier 
literature, that the family structure has 
a group theoretic origin, with all three families embedded in a large 
representation of a family unification group.  Since there are 15 (or if 
right handed neutrinos are included) 16 Weyl spinor fields in each 
family, a group representation of dimension at least 45 or 48 is required.   
So we are necessarily considering a large group representation, and if 
we invoke naturalness to require that it be a low-lying representation of  
its Lie group or algebra, then we are necessarily considering a large group.  
A particularly interesting candidate is the group $E_8$, 
which has a 248 dimensional Lie algebra and, as the largest exceptional  
group, a unique position in the standard Cartan classification of Lie groups.   
Our aim in this paper, which is unapologetically 
speculative and programmatic, is to review earlier work on $E_8$ unification,  
to explain difficulties encountered, and to argue that recent developments 
suggest that there may be mechanisms that can overcome these 
difficulties.  Thus  
the time may be ripe to reconsider $E_8$, and  
specifically $E_8$ supersymmetric Yang-Mills theory,  as a family 
unification model.  

Unification theories based on simple Lie groups follow a basic paradigm  
established by the $SU(5)$ and $SO(10)$ models.  The gauge bosons are 
as usual in  
the adjoint representation of the group, and left-handed Weyl fermions are 
placed in one or more additional representations, chosen to give 
cancellation of anomalies together with the standard model fermion structure  
under breaking of the unification group 
to $SU(3) \times SU(2) \times U(1)$.  Turning to $E_8$, this is the 
unique simple Lie group in which the adjoint 
representation, of dimension 248, is 
also the fundamental representation.  Hence the natural implementation of 
the basic paradigm is to place left-handed Weyl fermions in the 248 
representation, giving a model in which the gauge bosons or gluons, and 
the fermionic matter fields, are both in the adjoint 248 representation.  
Since in four dimensions supersymmetric Yang-Mills theory can be constructed 
with adjoint fermions that are either Majorana or Weyl [2], in this $E_8$ 
model the fermions and gluons are in the same supermultiplet, achieving a 
complete unification of matter fields and force-carrying fields.  The 
point that an $E_8$ unification model is automatically supersymmetric  
was made independently more than twenty years ago by Baaklini [3], 
by Bars and G\"unaydin [4], and by Konshtein and Fradkin [5], was 
followed up on in a paper of Koca [6], and was
briefly noted in 
Slansky's comprehensive review [7] of group theory for model building.  

Another interesting feature of $E_8$ is that it naturally contains three 
families.  Most of the recent discussions of single family  grand 
unification are based on either the group $SO(10)$ [8] or the group 
$E_6$ [9].  In $SO(10)$ unification the 16 Weyl fermions of a family 
(including a right handed neutrino)  are placed in a 16 representation, while 
in unification in the larger group $E_6$, of which $SO(10)$ is a subgroup, 
these fermions are placed in a 27 representation.   Under the decomposition 
$E_8 \supset SU(3) \times E_6$, the 248 of $E_8$ branches [7] as 
$$248=(8,1)+(1,78)+(3,27)+(\overline 3,\overline{27})~~~,\eqno(1a)$$
while under $E_8 \supset SU(3) \times SO(10) \times U(1)$, the 248 
branches [7] as 
$$\eqalign{
248=&(1,16)(3)+ (1,\overline{16})(-3) + (3,16)(-1)
+(\overline 3,\overline{16})(1)+(3,10)(2)\cr
+&(\overline 3,10)(-2)+(3,1)(-4)
+(\overline 3,1)(4)+(8,1)(0)+(1,45)(0)+(1,1)(0)~~~,\cr
}\eqno(1b)$$
with the $U(1)$ generator in parentheses.  Thus, the 248 of $E_8$ naturally 
contains three 27's of $E_6$ and three 16's of $SO(10)$, and so can unify 
the three families into a single representation.  The point that $E_8$ 
Yang-Mills theory 
can contain $SU(3)$ as a family group was made by Bars and G\"unaydin [4] 
and was emphasized in Slansky's review [7] and also by Barr [10].  
In the different dynamical 
context of supersymmetric nonlinear $\sigma$ models, the point that 
$E_8$ can naturally lead to three families was made in papers of Ong [11], 
Buchm\"uller and Napoly [12], Itoh, Kugo, and Kunitomo [13], and 
Ellwanger [14].  

Despite the attractive features of automatic supersymmetry and natural  
inclusion of three families, the reason that $E_8$ 
has not been further pursued as a unification group is that in addition 
to three families,  it contains three mirror families.  Thus, under 
$E_8 \supset SU(3) \times E_6$,  in addition to three 27's there are 
three $\overline {27}$'s, while under 
$E_8 \supset SU(3) \times SO(10) \times U(1)$, in addition to three 16's 
there are three $\overline{16}$'s.  The presence of mirror families 
leads to potential phenomenological and theoretical difficulties.  

The phenomenological difficulty is that since the masses of mirror 
families break $SU(2) \times U(1)$ electroweak symmetry, they must be 
of order the electroweak symmetry breaking scale, at most a few hundred 
GeV.  Hence, although they need not have been produced in current accelerator  
experiments, they will manifest themselves indirectly through electroweak  
radiative corrections, and should be copiously produced once the large hadron 
collider (LHC) is 
operative.  A detailed review of experimental signatures for mirror fermions 
has been given by Maalampi and Roos [15] (see also Montvay [16] and 
Triantaphyllou [17]; the latter also has discussed a possible role for 
mirror fermions in dynamical electroweak symmetry breaking.)   
One potential phenomenological objection 
to mirror fermions is that under the assumptions that their masses are 
much larger than the $Z$ boson 
mass and are degenerate within right-handed doublets, 
each family of mirror fermions would make a contribution of $2/(3 \pi)$ to 
the electroweak $S$ parameter (see Peskin and Takeuchi [18], and the review 
by Erler and Langacker [19]), in strong disagreement with experiment.   
However, this is not as definitive as it seems; when the degeneracy 
assumption is dropped the contribution of a mirror family to $S$ can have 
either sign (or be zero), and recent analyses of the electroweak precision 
data by Novikov, Okun, Rozanov, and Vysotsky [20] and by He, Polonsky, and 
Su [20] conclude that additional 
chiral generations are not currently excluded, with Novikov et al. finding 
a chi-squared minimum 
between one and two extra generations.   An second analysis by Choudhury, Tait 
and Wagner [21], focusing on  
additional mirror bottom quarks, also finds an improved fit to the 
electroweak data.  In an $E_8$ unification model, each 
fermion family is accompanied by a family of vector gluons which will also,  
at least [22] in the case of non-mass degenerate vector doublets, 
make contributions to the $S$ parameter, and therefore will further weaken 
the constraints coming from the electroweak data.  Thus the mirror structure 
predicted by an $E_8$ model may well be consistent with current data.  

The theoretical difficulty is that under the most attractive channel rule, 
a theory with equal numbers of ordinary and mirror families would in general 
be expected to form a chiral symmetry breaking family-mirror family 
condensate, and so one would naively expect {\it no} low energy families to 
survive in the low energy effective action.  This expectation has become   
virtual dogma in model building, where it is usually stated that in a 
model with $n_f$ families and $n_{\bar f}$ mirror families, the difference    
$n_f-n_{\bar f}$ gives the number of surviving low energy families if 
positive, and the number of surviving low energy mirror families if negative.  
However, this dogma must be treated with some skepticism, since there are 
known instances (see Seiberg [23] and Holdom and Roux [24]) where the 
most attractive channel rule breaks down.  In the specific context of 
supersymmetric $E_8$ Yang-Mills theory, the issue is whether an $E_8$ 
singlet gluino condensate $\langle \lambda \lambda \rangle$ forms, as 
suggested by an effective action argument of Veneziano and Yankielowicz [25].  
The presence of such a condensate would prevent the appearance of fermions 
(which are the $E_8$ gluinos) in the low-energy effective action.  

Recently, in a very interesting paper, Kovner and Shifman [26] 
have argued that the 
Veneziano-Yankielowicz effective action must be modified so as to explicitly 
exhibit the $Z_{2T(G)}$ discrete chiral symmetry, which is the nonanomalous 
remnant of the anomalous $U(1)$ axial symmetry generated by phase rotations 
of the gluino fields. (Here  $\ell=2T(g)$  is the Dynkin index of the adjoint  
representation, which equals 60 for the adjoint 248 representation of $E_8$.)   
They show that there is a simple modification of the Veneziano-Yankielowicz 
action which has the required discrete symmetry, and that this action 
predicts that there is a phase in which the discrete chiral symmetry is 
unbroken, and thus in which the usual singlet gluino condensate does not 
develop.  While the two independent arguments advanced by Kovner and Shifman 
to support their suggestion for a new phase are now 
discounted (one of these was 
based on problems with the Witten index for certain groups, which are now 
resolved [27]; the other on a mismatch between the strong and weak coupling 
instanton calculations of the gluino condensate, which has been given 
another explanation [28]), their effective action argument for the existence 
of a phase without a gluino condensate is still viable, and their conjecture 
of a new phase for supersymmetric gluodynamics is open, although still  
debated [29, 30].  

In particular, although Cs\'aki and Murayama [30] have used discrete anomaly 
matching to argue against the Kovner-Shifman vacuum, their argument  
assumes that the ground state spectrum consists of hypercolor (here $E_8$ )
singlets.  Thus it does not rule out the possibility that the Kovner-Shifman  
vacuum is in a trivial, deconfined phase with the same particle spectrum 
as the starting $E_8$ gauge theory, before symmetry breaking 
arising from perturbations to the SUSY gluodynamics structure is taken  
into account.  A deconfined phase would obey all anomaly matching 
constraints, and even if not generic for SUSY Yang-Mills gluodynamics, its  
presence just in special cases including $E_8$ would suffice for the 
arguments we are making.     

If supersymmetric Yang-Mills for the $E_8$ group is in the Kovner-Shifman 
vacuum, then the principal theoretical objection to $E_8$ as a unification 
group disappears, since the theory in isolation would remain a supersymmetric 
theory (as assured by Witten index arguments [27,31]) with massless gluinos    
in the Kovner-Shifman phase.   
Of course, to get a realistic theory breaking of both $E_8$ 
symmetry and supersymmetry is needed.   
As noted by Shifman and Vainshtain [29] 
(in the course of a discussion of the Witten index, but their remark is 
more generally  relevant) the Kovner-Shifman vacuum is ``potentially 
unstable under various deformations.''    One obvious deformation that 
could be relevant is the embedding of supersymmetric $E_8$ in supergravity. 
When the gravitino and graviton fields are integrated out at tree level, 
one obtains [32] a supersymmetric four-gluino effective action that could 
be the trigger for dynamical symmetry breaking of either or both the 
$E_8$ internal symmetry and supersymmetry.  Supersymmetry breaking could 
also arise from supersymmetry breaking in another sector of the theory  
(such as the second $E_8$ expected in string theory) communicated by the  
supergravity interaction between the two; a general review of this approach 
is given in Weinberg [33], and an application of gravity-mediated 
supersymmetry breaking to 
sequential breaking of $E_8$ to $E_6$ and then to $SO(10)$ is 
discussed by Mahapatra and Deo [34]. 

As noted by Bars and G\"unaydin [4], an $E_8$ unification theory cannot have 
elementary Higgs scalars without losing the property of asymptotic freedom,   
because the Dynkin index of the smallest candidate Higgs representation 
(the 3875) is already too large.  Hence in an asymptotically free $E_8$ 
theory, all symmetry breaking (other than that 
communicated by gravity mediation from another sector) 
must be dynamical, through the formation 
of suitable condensates of the gluinos (and of gluinos and 
gluons as well, if condensate 
formation preserves supersymmetry).  Chiral symmetry breaking 
by condensate formation was reviewed some time ago by Peskin [35], and 
recently there has been much interest in the role of 
non-singlet condensates that break gauge symmetry, 
in the context of ``color superconductivity'' in high density QCD [36].  
In order to give the mirror fermions larger masses than the top quark, 
there must be a condensate that introduces an asymmetry between the fermions 
and their mirror partners.  One candidate arises from the fact that in 
$SU(3) \times E_6$ 
one has $(\overline{3}, \overline {27}) \times (\overline {3}, \overline{27}) 
\supset (6_s,27_s)$.  Since 
under the decomposition 
$E_6 \supset SO(10) \times U(1)$ one has $27 \supset 1(4)$,  
a gluino-gluino condensate with nonvanishing vacuum expectation of 
the $1(4)$ would preserve $SO(10)$ symmetry, while breaking the $U(1)$ 
factor and introducing an asymmetry between the three fermion families and 
their mirror families. Moreover, since under the family group decomposition  
$SU(3) \supset SU(2) \times U(1)$ the $6_s$ contains a singlet of $SU(2)$, 
this expectation would split two degenerate families apart from a third, 
approximating what is observed.  (From the viewpoint of $E_8$, the 
condensate we are proposing is contained in $3875_s \subset 
248 \times 248$, which is the second most attractive symmetric channel  
according to 
the most attractive channel rule.)
To break $SO(10)$ down to the 
standard model further 
condensates would be needed; we note that all of the Higgs 
representations used 
in models for the breaking of $SO(10)$ unification are contained in the 
$248 \times 248$ of $E_8$, and so could be generated by the formation of 
non-singlet gluino-gluino condensates.   As a final remark on symmetry 
breaking, we mention that a much 
studied alternative to dynamical generation of Higgs condensates 
is their generation by dimensional reduction from a higher dimensional 
gauge theory; for a recent discussion of this mechanism as applied to 
$E_8$ and three family unification, in the context 
dimensional reduction over coset spaces, see Manousselis and Zoupanos [37].  

The phenomenology of a supersymmetric $E_8$-based grand unification and 
family unification model will differ significantly from that 
expected in the minimal supersymmetric standard 
model (MSSM) and its extensions.  As in the MSSM,
the superpartners for the gauge bosons 
in the $E_8$ model are spin-1/2 fermions, and $R$-parity conservation [38]   
implies that the lightest superpartner will be stable.  
However, in the $E_8$ theory, in addition to there being mirror fermions, 
the superpartners for the quarks and 
leptons are vectors rather than scalars.  Thus there are   
potentially observable signatures for $E_8$ unification  
at the LHC and other future facilities.

\bigskip
\centerline{\bf Acknowledgments}
This work was supported in part by the Department of Energy under
Grant \#DE--FG02--90ER40542.  I wish to thank Csaba Cs\'aki, 
Maurice Goldhaber, Paul Langacker, Michael 
Peskin, Nathan Seiberg, Jacob Sonnenschein, 
Edward Witten, and Ren-Jie Zhang for 
helpful discussions or email correspondence.  
\vfill\eject
\centerline{\bf References}
\bigskip 
\noindent
\item{[1]}  Three prominent types of ideas that have been advanced to  
explain family structure are (1)  composite models, in which the quarks 
and leptons are not elementary, and the different 
families are different internal symmetry states of their constituents,  
(2) string-inspired models, in which the different families correspond to 
string excitations with 
different topological quantum numbers of the background manifold on which 
the string spacetime is compactified, and (3) as discussed here, 
group theoretic family 
unification models, in which the quarks and leptons are elementary (at least 
below the Planck scale) and the family structure arises because the 
fermions lie in a large group representation. For early discussions 
of family groups, see P. Ramond, `The Family Group in Grand Unified 
Theories', invited talk at the Sanibel Symposia, 1979 (reissued as 
ArXiv:  hep-ph/9809459); and A. Zee, 
{\it Unity of Forces in the Universe} Vol. I (World Scientific, 1982),   
Sec. VIII.  A pedagogical account of composite and string-inspired models 
is given in Chapt. 8 and Chapt. 17, respectively, of R. N. Mohapatra, 
{\it Unification and Supersymmetry}, Second Ed. (Springer, 1992). 
For a recent discussion of empirical systematics of the family problem, 
and further references, see M. Goldhaber,  Proc. Nat. Acad. 
Sci. USA 99 (2002) 33 and arXiv: hep-ph/0201208.  
\bigskip
\noindent
\item{[2]}  L. Brink, J. H. Schwarz and J. Scherk, Nucl. Phys. B121 (1977), 
77, show that  the Fierz identity needed to verify supersymmetry in four 
dimensional Yang-Mills theory holds for Weyl as well as Majorana fermions. 
One can also directly transform supersymmetric Yang-Mills theory from 
Majorana to Weyl fermions, using the Pauli-G\"ursey transformation (which is 
a special instance of a Bogoliubov transformation); see 
W. Pauli, Nuovo Cimento 6 (1957) 204 and F. G\"ursey, Nuovo Cimento 7 
(1958) 411; R. E. Marshak, Riazuddin and C. P. Ryan, {\it Theory 
of Weak Interactions in Particle Physics} (Wiley-Interscience, 1969), 
pp. 71-72.  
\bigskip
\noindent
\item{[3]}  N. S. Baaklini, Phys. Lett.  91B (1980) 376.
\bigskip 
\noindent
\item{[4]}  I. Bars and M. G\"unaydin, Phys. Rev. Lett.  45 (1980) 859. 
\bigskip
\noindent
\item{[5]}  S. E. Konshtein and E. S. Fradkin, Pis'ma Zh. Eksp. Teor. Fiz. 
32 (1980) 575 [English translation:  JETP Lett.  32 (1981) 557] .
\bigskip
\noindent
\item{[6]} M. Koca, Phys. Lett. 107B (1981) 73. 
\bigskip
\noindent
\item{[7]}  R. Slansky, Phys. Reports {\bf 79} (1981) 1.  
\bigskip
\noindent
\item{[8]}  For an introduction to $SO(10)$ unification and reprints of 
seminal papers, see A. Zee, Ref. [1], Sec. IV.  
For a more detailed pedagogical account, 
see R. N. Mohapatra, Ref. [1], Chapt. 7. 
\bigskip
\noindent
\item{[9]}  For an introduction to $E_6$ unification and reprints of 
seminal papers, see A. Zee, Ref. [1], Sec. V.  
For a review of $E_6$ phenomenology, see J. L. Hewett 
ad T. G. Rizzo, Phys. Rep. 183 (1989) 193.  
\bigskip
\noindent
\item{[10]}  S. M. Barr, Phys. Rev. D37 (1988) 204.
\bigskip
\noindent
\item{[11]}  C. L. Ong, Phys. Rev. D31 (1985) 3271.
\bigskip
\noindent
\item{[12]}  W. Buchm\"uller and O. Napoly, Phys. Lett. 163B (1985) 161.
\bigskip
\noindent
\item{[13]}  K. Itoh, T. Kugo and H. Kunitomo, Prog. Theor. Phys. 75 (1986) 
386.
\bigskip
\noindent
\item{[14]} U. Ellwanger, Nucl. Phys. B356 (1991) 46.   
\bigskip     
\noindent
\item{[15]}   J. Maalampi and M. Roos, Phys. Reports 186 (1990) 53.
\bigskip
\noindent 
\item{[16]}   I. Montvay, Phys. Lett. 205B, 315 (1988).
\bigskip
\noindent
\item{[17]}  G. Triantaphyllou, Int. Journ. Mod. Phys. A (2000) 265;  
J. Phys. G:  Nucl. Part. Phys. 26 (2000) 99.  These papers focus on the 
phenomenology of mirrors.  For a discussion of dynamical symmetry breaking 
by mirrors with new strong interactions, see G. Triantaphyllou, 
`Dynamical Symmetry Breaking with Mirror 
Fermions', talk at the Corfu Summer Institute on Elementary Particle 
Physics, 1998, arXiv: hep-ph/9908251; Mod. Phys. Lett. A16 (2001) 53; 
`The Weak Scale:  Dynamical Determination Versus Accidental Stabilization', 
talk at the Corfu Summer Institute on Elementary Particle Physics, 2001, 
arXiv:  hep-ph/0109023.  
\bigskip
\noindent
\item{[18]}  M. E. Peskin and T. Takeuchi, Phys. Rev. D 46 (1992) 381.  
\bigskip
\noindent
\item{[19]}  J. Erler and P. Langacker, `Electroweak Model and Constraints 
on New Physics', in Particle Data Group, 
{\it Review of Particle Physics}, 2000, 2001 (available online at   
http://pdg.lbl.gov/).
\bigskip
\noindent
\item{[20]}  V. A. Novikov, L. B. Okun, A. N. Rozanov and M. I. Vysotsky, 
`Extra generations and discrepancies of electroweak precision data', 
arXiv: hep-ph/0111028; H.-J. He, N. Polonsky, and S. Su, `Extra Families, 
Higgs Spectrum and Oblique Corrections', arXiv: hep-ph/0102144.  
\bigskip
\noindent
\item{[21]}  D. Choudhury, T. M. P. Tait and C. E. M. Wagner, `Beautiful 
Mirrors and Precision Electroweak Data', arXiv: hep-ph/0109097. 
\bigskip
\noindent
\item{[22]}  R.-J. Zhang and S. L. Adler, unpublished.  Degenerate vector 
doublets make a vanishing contribution to the $S$ parameter, but 
nondegenerate doublets should contribute a logarithm similar in structure,  
but with a different numerical coefficient that has not been computed,  
to that found in the fermion case.  
\bigskip
\noindent
\item{[23]} N. Seiberg, `The Power of Holomorphy - Exact Results in 
4D SUSY Field Theories', arXiv: hep-th/9408013, Secs. 3 and 5.  See also 
K. Intriligator and N. Seiberg, `Lectures on Supersymmetric 
Gauge Theories and Electric-Magnetic Duality', arXiv: hep-th/9509066, 
Secs. 4.3 and 
5.4; to appear in the Proceedings of the Trieste '95 spring school, 
TASI '95, Trieste '95 summer school, and Cargese '95 summer school.  
\bigskip
\noindent
\item{[24]}  B. Holdom and F. S. Roux, Phys. Rev. D59 (1999) 015006.  
\bigskip
\noindent
\item{[25]}  G. Veneziano and S. Yankielowicz, Phys. Lett. 113B (1982) 231;  
see also T. R. Taylor, G. Veneziano and S. Yankielowicz, Nucl. Phys.  B218 
(1983) 493; H. P. Nilles, Phys. Lett. 112B (1982) 455 and Phys. Lett. 115B   
(1982) 193.  
\bigskip
\noindent
\item{[26]} A. Kovner and M. Shifman, Phys. Rev. D56 (1997) 2396.
\bigskip
\noindent    
\item{[27]}  E. Witten, `Supersymmetric Index in Four-Dimensional Gauge 
Theories', arXiv: hep-th/0006010.
\bigskip
\noindent
\item{[28]} T. J. Hollowood, V. V. Khoze, W. Lee and M. P. Mattis, Nucl. 
Phys. B 570 (2000) 241.
\bigskip
\noindent
\item{[29]}  I. I. Kogan, A. Kovner and M. Shifman, Phys. Rev. D57 (1998) 
5195; M. Shifman and A. Vainshtein, `Instantons Versus Supersymmetry: 
Fifteen Years Later', arXiv: hep-th/9902018; A. Ritz and A. Vainshtein, 
Nucl. Phys. B 566 (2000) 311; Y. Shadmi and Y. Shirman, Rev. Mod. Phys. 
72 (2000) 25, note 11 on p. 33; V. S. Kaplunovsky, J. Sonnenschein, and 
S. Yankielowicz, Nucl. Phys. B552 (1999) 209.  
\bigskip
\noindent
\item{[30]}  C. Cs\'aki and  H. Murayama, Nucl. Phys. B515 (1998) 114. 
\bigskip
\noindent
\item{[31]}  For a survey of the Witten index analysis of supersymmetric 
gauge theories, see Secs. 29.1 and 29.4 of 
 S. Weinberg {\it The Quantum Theory of 
 Fields}, Vol. III: Supersymmetry (Cambridge University Press, 2000).
\bigskip
\noindent
\item{[32]}  S. L. Adler, Ann. Phys. 290 (11) 2001.
\bigskip
\noindent
\item{[33]}  For a survey of gravity mediated supersymmetry breaking, 
see Sec. 31.7 of Weinberg, Ref. [31].  
\bigskip
\noindent
\item{[34]}  S. Mahapatra and B. B. Deo, Phys. Rev. D38 (1988) 3554.
\bigskip
\noindent
\item{[35]}  M. E. Peskin, `Chiral Symmetry and Chiral Symmetry Breaking', 
in {\it Recent Advances in Field Theory and Statistical Mechanics} 
(Les Houches,  
1982), J.-B. Zuber and R. Stora, eds. (North-Holland, 1984).  
\bigskip
\noindent
\item{[36]}  For a recent review, see M. Alford, `Color Superconducting 
Quark Matter', in {\it Annual Review of Nuclear and Particle Science}, 
V. 51 (Annual Reviews, 2001). 
\bigskip
\noindent
\item{[37]}  P. Manousselis and G. Zoupanos, `Dimensional Reduction over 
Coset Spaces and Supersymmetry Breaking', arXiv: hep-ph/0111125.
\bigskip
\noindent
\item{[38]}  The $R$-parity is defined by $\Pi_R=(-1)^F(-1)^{3(B-L)}$, 
with $(-1)^F$ the fermion parity which is $+1$ for all bosons 
(irrespective of 
whether vector or scalar) and $-1$ for all fermions.  Hence in the $E_8$ 
theory, as in the MSSM, all standard model particles have $R=1$ and all 
superpartners have $R=-1$.  For further details and references, see 
Weinberg, Ref. [31], p. 184 and Mohapatra, Ref. [1], p. 277.

\vfill
\eject
\bye